\documentclass[journal=jpccfk,manuscript=article]{achemso}

\usepackage{chemformula} 
\usepackage[T1]{fontenc} 
\usepackage{amsmath, amssymb}

\author{Kunyang Sun}
\affiliation{Department of Chemistry and Biochemistry, UC San Diego, La Jolla, CA, 92093, USA}
\alsoaffiliation{Pitzer Center for Theoretical Chemistry and Department of Chemistry, University of California, Berkeley, CA, 94720, USA}
\author{Matthew Du}
\email{madu@uchicago.edu}
\affiliation{Department of Chemistry and Biochemistry, UC San Diego, La Jolla, CA, 92093, USA}
\alsoaffiliation{Department of Chemistry and The James Franck Institute, University of Chicago, Chicago, Illinois, 60637, USA}
\author{Joel Yuen-Zhou}
\email{joelyuen@ucsd.edu}
\affiliation{Department of Chemistry and Biochemistry, UC San Diego, La Jolla, CA, 92093, USA}

\title{Exploring the Delocalization of Dark States in a Multimode Optical Cavity}

\begin{document}

\begin{tocentry}




\includegraphics{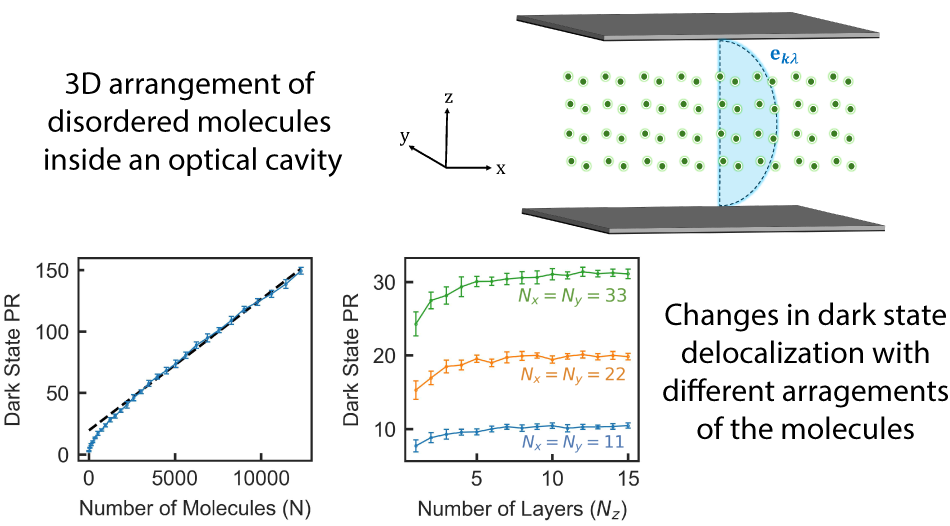}
\end{tocentry}

\begin{abstract}
    The strong coupling between molecules and photonic modes in a Fabry-P\'{e}rot optical cavity, which forms hybrid light-matter states called polaritons, has been demonstrated as a promising route to control the rates of chemical reactions.
    However, theoretical studies, which largely employ models with a single cavity mode, cannot explain the experimentally observed kinetic changes.
    While simplified multimode models with one spatial dimension can capture experimental features involving the polariton states, it is unclear whether they can also describe the dark states.
    Here, we study the delocalization of dark states for molecules in a multimode cavity, accounting for 
    the three-dimensional nature of experimental setups.
    Accounting for energetic and orientational disorder, but fixing Rabi splitting and intermolecular distances (i.e., no positional disorder), we find that the delocalization of the dark states scales linearly with the number of molecules in the plane parallel to the cavity mirrors, in contrast to one-dimensional multimode models. 
    Adding layers of molecules along the axis normal to the mirrors increases the delocalization much less.
    Similar to the one-dimensional models, the dark-state delocalization is enhanced for smaller values of molecular energetic disorder, relative to the light-matter coupling, and cavities with longer longitudinal length.
    Our work indicates that for certain phenomena, understanding the dark states under strong light-matter coupling might require a proper multimode description of the optical cavity. 
\end{abstract}

\section{Introduction}
Optical cavities have demonstrated potential in controlling chemical reactivity and other molecular properties \cite{Ebbesen2016,Ribeiro2018,Feist2018,Flick2017,Flick2018,Hertzog2019,Hirai2020rev,Xiang2021, Simpkins2021, Dunkelberger2022,Mandal2023,Wang2021,Gu2023,George2023}. 
Typically, a three-dimensional (3D) arrangement of many molecules is placed in a Fabry-P\'{e}rot cavity~\cite{Ebbesen2016, Simpkins2023}, where light is confined between two highly reflective planar mirrors.
The photonic modes propagate freely along two spatial dimensions and consist of two polarizations per wavevector.
If the collective light-matter coupling is sufficiently strong, an ensemble of molecular excitations and a (nearly) resonant photonic mode can coherently exchange energy multiple times before the light escapes the cavity.
In this regime, the molecular and photonic states hybridize into states known as polaritons~\cite{Torma2014,Ebbesen2016,Baranov2018}, which differ in energy from their uncoupled constituents. 
The formation of polariton states has been found to correlate with modified rates of thermally activated chemical reactions~\cite{Thomas2016, Thomas2019, Lather2019, Hirai2020, Hirai2020rev, Ahn2023}, suggesting a direct influence on reaction dynamics.

However, existing theoretical models have not been able to explain the experimental kinetic changes in a unified manner~\cite{Wang2021, Li2022, Campos-Gonzalez-Angulo2023, Mandal2023}.
Models involving a single molecule predict modified reaction rates~\cite{Galego2019, Li2021natcomm, Li2021jpcl, Yang2021, Schafer2022, Lindoy2022, Mondal2022, Philbin2022, Lindoy2023, Lindoy2024}, but the assumed light-matter coupling per molecule is significantly larger than experimental values. 
Conversely, models with many molecules predict negligible changes in kinetics under standard experimental conditions~\cite{Galego2019, Campos-Gonzalez-Angulo2020, Zhdanov2020, Li2020, Vurgaftman2020, Du2021, Wang2022, Du2023, Lindoy2024, Pannir-Sivajothi2025}. 
Indeed, conventional transition-state theory yields no rate modification in the thermodynamic limit of molecules~\cite{Galego2019, Campos-Gonzalez-Angulo2020, Zhdanov2020, Li2020}, since the polariton states are vastly outnumbered by the optically dark molecular states, whose energy distribution is essentially unchanged by strong light-matter coupling. 
Only under special conditions have theories predicted altered reactivity in the thermodynamic limit~\cite{Galego2019, Campos-Gonzalez-Angulo2019, Du2021, Yang2021, Du2022, Wang2022, Mandal2022, Lindoy2024, Vega2024, Pannir-Sivajothi2025}.
In most many-molecule models, which are based on the Tavis-Cummings model \cite{Tavis1968}, there is just 1 photonic mode.
Consequently, these models also cannot capture why the reactivity is altered only when the molecular states are (nearly) resonant with cavity modes that have zero in-plane wavevector~\cite{Thomas2016, Thomas2019, Lather2019, Hirai2020, Hirai2020rev, Ahn2023}. 

Various properties of molecular polaritons can be well understood by models with one spatial dimension along which $N$ molecules are distributed and $N$ photonic modes can propagate freely.
For each (in-plane) wavevector, only one polarization of light is considered.
This simplified representation of the experimental setup has been successful in modeling the relaxation dynamics of polaritons~\cite{Agranovich2003, Michetti2008, Michetti2009, Tichauer2021}, agreeing with experimental observations~\cite{Lidzey2002, Coles2013}.
Similarly, such a model~\cite{Sokolovskii2023} has been applied to explain experiments showing polariton propagation~\cite{Rozenman2018, Pandya2021}. 
It has also been demonstrated that the energetic disorder of the molecules localizes certain polariton states~\cite{Michetti2005, Agranovich2007, Michetti2008, Litinskaya2006, Litinskaya2008, Engelhardt2023}, where some of these states eventually enter a turnover regime and actually become less localized as disorder is further increased~\cite{Engelhardt2023}. 
In addition, strong coupling affects molecular properties the most when the molecular states have higher energy than a cavity mode with zero in-plane wavevector~\cite{Ribeiro2022}, in contrast to models with 1 cavity mode.

However, the 1D multimode models might be too simple to capture the properties of the dark states, in particular, their delocalization.
The delocalization of dark states under strong coupling has been theoretically shown to impact processes such as energy transport~\cite{Botzung2020} and chemical reactions~\cite{Du2022}.
In 1D multimode models without direct intermolecular coupling, the dark states can be delocalized across 3-4 molecules~\cite{Ribeiro2022}.
This result is similar to models with a single cavity mode, which yield a dark-state delocalization ``length" of 2-3 molecules~\cite{Botzung2020, Du2022} (see Ref~\cite{Liu2025} for related findings on polariton delocalization).
For both types of models, as the total number of molecules increases, the delocalization of the dark states approaches the aforementioned values.
This convergence of delocalization appears in stark contrast to a finding of Bradbury et al.~\cite{Bradbury2024}, who theoretically study a two-dimensional layer of molecules in a cavity.
As in experiments, the photonic modes propagate freely along two dimensions and consist of multiple polarizations per wavevector. 
The authors find that the dark states can be delocalized across hundreds of molecules~\cite{Bradbury2024}, suggesting that the extent of delocalization might not converge in the thermodynamic limit of molecules.

\begin{figure}
    \centering
    \includegraphics[width = 1.0 \textwidth]{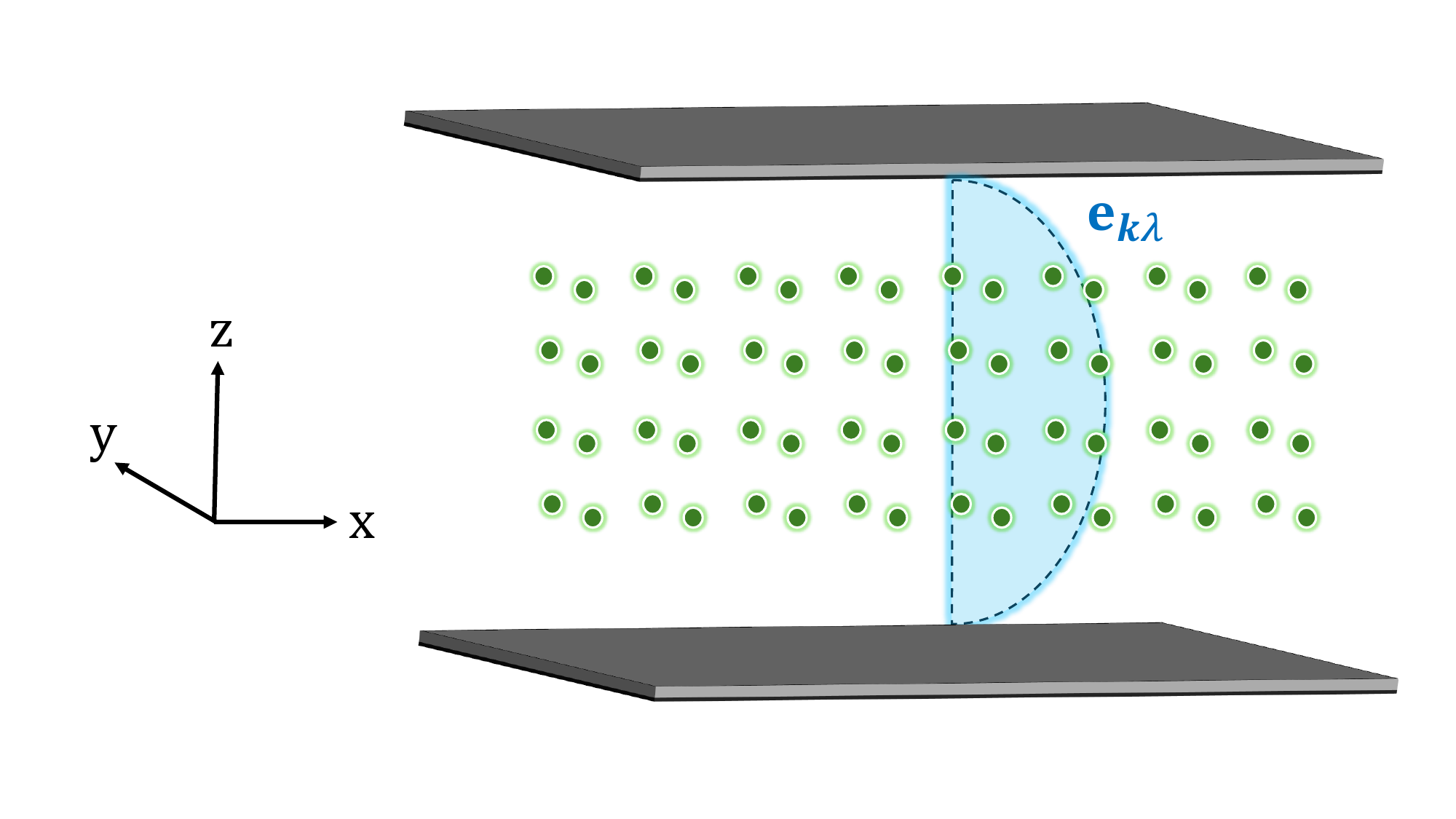}
    \vspace{-40pt}
    \caption{\textbf{Schematic representation of the setup considered in this work.} A three-dimensional array of molecules (green dots) interacts with photonic modes, defined by in-plane wavevector $\mathbf{k}$ and polarization $\lambda$, inside a Fabry-P\'{e}rot cavity.
    The electric field amplitude [given by polarization vector $\mathbf{e}_{\mathbf{k}\lambda}$; see Eq.~\eqref{eq:e_k}] for photonic mode $(\mathbf{k}, \lambda)$ is illustrated in blue.}
    \label{fig:cavity}
\end{figure}

Here, we investigate the delocalization of the dark states for molecules in a Fabry-P\'{e}rot cavity. As shown in Figure \ref{fig:cavity}, our theoretical model captures aspects of typical experimental settings by including multiple photon modes, multiple polarizations per wavevector, and a three-dimensional arrangement of molecules with energetic and orientational disorder.  
We explored various design choices of the system to see how they would change the dark-state delocalization. 
We find that as the number of molecules increases, the dark states become more delocalized, where the extent of delocalization scales linearly for sufficiently large systems, in contrast to single-mode and 1D multimode models. 
As we add more layers of molecules along the longitudinal axis of the cavity ($z$-axis, Figure~\ref{fig:cavity}), the delocalization also increases but to a lesser degree, converging at a value comparable to the single-layer value. 
However, when we perturb the degree of molecular energy disorder and the cavity length, the dark-state delocalization changes similarly compared to a one-dimensional model~\cite{Ribeiro2022}. 
Overall, the results here shed light on the properties of the dark states for an experimentally representative model of molecules in an optical cavity.

\section{Model and Methods}
\subsection{Model Hamiltonian}
\label{sec:hamiltonian}
For the optical cavity, we consider two planar mirrors separated by a distance of $L_z$ in the $z$ direction and assume periodic boundary conditions in the $x$ and $y$ directions (Figure~\ref{fig:cavity}). 
The cavity has a volume $V = L_xL_yL_z$, where $L_w$ is the length of the cavity along the $w = x,y,z$ axis.
Along the $z$ axis (confinement or longitudinal axis), we define the boundaries of the cavity to be at $z=0$ and $z=L_z$. 
To include as many molecules as possible in our calculations below, here we assume only the lowest band of the photon modes. Within this band, each cavity mode is characterized by the wavevector $\mathbf{k} = (2\pi m_x/L_x, 2\pi m_y/L_y)$ for $m_x, m_y \in \mathbb{Z}$ and polarization $\lambda = s,p$. The cavity mode $(\mathbf{k}, \lambda)$ has a polarization-independent frequency $\omega_{c, \mathbf{k}} = (c/\sqrt{\epsilon}) \sqrt{k^2 + (\pi/L_z)^2}$, where $c$ is the speed of light in vacuum, $\epsilon$ is the dielectric constant inside the cavity, and $k=|\mathbf{k}|$. 
The polarization vector associated with the mode $(\mathbf{k}, \lambda)$ is~\cite{Zoubi2005}
\begin{equation}
\mathbf{e}_{\mathbf{k}\lambda}(z) = 
\begin{cases} 
i \sin\left(\frac{\pi z}{L_z}\right) (\hat{k} \times \hat{z}), & \lambda = s, \\
\frac{c\pi}{\sqrt{\epsilon} L_z \omega_{c,k}} \left[ \sin\left(\frac{\pi z}{L_z}\right) \hat{k} - i\frac{k L_z}{\pi} \cos\left(\frac{\pi z}{L_z}\right) \hat{z} \right], & \lambda = p
\end{cases}
\label{eq:e_k}
\end{equation}
for $k \neq 0$. For $k = 0$, we can choose any orthogonal vectors in the $xy$ plane for the two different polarizations. Thus, we conveniently set
\begin{equation}
\mathbf{e}_{(k=0)\lambda}(z) = 
\begin{cases}
i \sin\left(\frac{\pi z}{L_z}\right) \hat{y}, & \lambda = s, \\
\sin\left(\frac{\pi z}{L_z}\right) \hat{x}, & \lambda = p.
\end{cases}
\end{equation}

Regarding the molecules, we take each to have exactly one electronic excited state. We assume that there is static disorder in both the transition energies and the transition-dipole-moment orientations. We place $N = N_x N_y N_z$ molecules inside the cavity such that they form a finite lattice whose crystal axes are $x$, $y$, and $z$ and whose corresponding dimensions are $N_x$, $N_y$, and $N_z$, respectively. Along each axis, we take the molecules to be evenly distributed by a fixed spacing $a_w$ for $w = x, y, z$. More specifically, we ensure that the average of the molecular coordinates will be 0 along the $x$ and $y$ axes and $L_z/2$ along the $z$ axis. Along the $x$ and $y$ axis, we modify the length of the cavity to just be able to contain all molecules in those dimensions with $L_w = a_w (N_w - 1)$ for $w = x, y$. Since the cavity length ($L_z$) determines the energy of the photon modes, when placing molecules along the $z$-axis, we do not modify the width of the cavity. Instead, we first stack the layers with the fixed spacing and then align the center of the stacked layers with the center (antinode) of the cavity mirrors (Figure~\ref{fig:cavity}). In other words, the molecules have positions $\mathbf{r} = (n_x a_x - L_x/ 2, n_y a_y - L_y / 2, L_z/2 + (1/2 - N_z/2 + n_z) a_z)$, where $n_w = 0, 1, \ldots, N_w - 1$ and $a_w$ is a fixed spacing for $w = x, y, z$. In our calculations, including large positional disorder (i.e., randomly displacing each molecule by up to 50\% of the lattice spacing along each in-plane dimension $w = x,y$ according to a uniform distribution) did not show any observable impact on dark-state participation ratio, in agreement with previous work~\cite{Michetti2005}. Therefore, for the remainder of this work, we have disregarded positional disorder, which remains a valid approach for small deviations from a perfect square lattice in the molecular arrangement.

For the molecule at position $\mathbf{r}$, the excited state is represented by $|e_\mathbf{r}\rangle$, which has an excited-state energy $\hbar \omega_{e,\mathbf{r}}$ and a transition dipole moment $\mu_\mathbf{r} = (\mu, \theta_\mathbf{r}, \phi_\mathbf{r})$ in spherical coordinates. To model the effects of diagonal energetic disorder, $\omega_{e,\mathbf{r}}$ is sampled from a normal distribution with a mean of $\hbar \omega_e$ and a standard deviation of $\sigma_e$. For off-diagonal orientational disorder, the angles $\theta_\mathbf{r} \in [0, \pi]$ and $\phi_\mathbf{r} \in [0, 2\pi)$ are chosen based on probability density functions $f_\theta(\theta_\mathbf{r}) = \frac{1}{2} \sin \theta_\mathbf{r}$ and $f_\phi(\phi_\mathbf{r}) = \frac{1}{2\pi}$, respectively.
We treat all disorder as uncorrelated between sites.
We have tried turning off orientational disorder and found appreciable but essentially quantitative changes to results involving delocalization, in agreement with previous works on 1D multimode models~\cite{Michetti2005, Ribeiro2022}.
Therefore, energetic and orientational disorder both matter, although the former affects the qualitative findings more than the latter.

For a system with $N$ molecules, we then have $2N$ photonic modes considering both polarization (we ignore contributions from higher order Brillouin zones to make calculations with large $N$ computationally feasible). We employ the rotating-wave approximation and the Coulomb gauge for the light-matter coupling terms. Therefore, assuming the system is in the single-excitation manifold, the total Hamiltonian \( H \) for a disordered molecular ensemble in a multimode optical cavity is expressed as:
\begin{align}
    H & = \sum_\mathbf{r} \hbar \omega_{e,\mathbf{r}} |e_\mathbf{r} \rangle \langle e_\mathbf{r}| + \sum_{\mathbf{k},\lambda} \hbar \omega_{c,\mathbf{k}} |c_{\mathbf{k}\lambda} \rangle \langle c_{\mathbf{k}\lambda}| \notag \\
    & - \sum_{\mathbf{r},\mathbf{k},\lambda} \mu \sqrt{\frac{\hbar \omega_{c,\mathbf{k}}}{2\epsilon V}} \left[ e^{i\mathbf{k} \cdot \mathbf{r}} (\hat{\mu}_\mathbf{r} \cdot \mathbf{e}_{\mathbf{k}\lambda}(z)) |e_\mathbf{r} \rangle \langle c_{\mathbf{k}\lambda}| + \text{h.c.} \right],
\end{align}
where the state $|c_{\mathbf{k}\lambda}\rangle$ represents the excitation in cavity mode $(\mathbf{k},\lambda)$, and h.c. stands for Hermitian conjugate.
For $N_x, N_y \gg 1$, the Rabi splitting at resonance (i.e., $\omega_{c,\mathbf{k}}=\omega_e$) is well approximated by $
\Omega_0 = 2\mu\eta \sqrt{\frac{N\hbar \omega_e}{2\epsilon V}}
$ where $ 
\eta = \sqrt{\frac{1}{3} \left\langle \sin^2\left(\frac{\pi z}{L_z}\right)\right\rangle_z}$
is a factor that accounts for the random orientation of the molecules, and $\langle \rangle_z$ denotes an average over the z-coordinates of the molecules.
Throughout this work, we choose values of energetic disorder that are small enough to satisfy the rotating-wave approximation, i.e., the Rabi splitting is less than 10\% of the energies of doubly excited states containing one excitation in the resonant photon mode and one excitation in a molecule~\cite{Frisk_Kockum2019, Forn-Diaz2019}.

\subsection{Dark-State Participation Ratio}

Since we are interested in exploring how the dark state localization is affected when coupling to a three-dimensional cavity with multiple photon modes, a natural way of describing such an effect is the participation ratio (PR). Solving the Hamiltonian above, we can obtain eigenvectors in the following form:
\begin{equation}
    |\psi\rangle = \sum_\mathbf{r} a(\mathbf{r}) |e_\mathbf{r}\rangle + \sum_{\mathbf{k}, \lambda} b(\mathbf{k}\lambda) |c_{\mathbf{k}\lambda}\rangle
\end{equation}
Here, following the definition in~\cite{Tichauer2021}, we define eigenstates containing less than 5\% of photon fractions as dark states. Therefore, the PR for dark states can be expressed as follows:
\begin{equation}
    PR_{dark} = \frac{(\sum_\mathbf{r} |a(\mathbf{r})|^2)^2}{\sum_\mathbf{r} |a(\mathbf{r})|^4}
\end{equation}
Consequently, \(PR_{dark}\) allows us to determine the degree of delocalization of each eigenstate across all molecular states.
Previous theoretical work~\cite{Du2022} by some of us has shown that $PR_{dark}$ can be a good proxy for changes in reaction rate due to strong coupling-induced delocalization of the dark states.
Below, we study $PR_{dark}$ averaged over all dark states and 10 disorder realizations.
We note that a quantity which is more directly related to dynamics and has similar behavior to the PR is the coherence length~\cite{Engelhardt2023}; a future comparison of the results in the current work with analogous studies focused on coherence lengths could be useful.


\section{Results and Discussion}
\begin{figure}
    \centering
    \includegraphics[width = 1.0\textwidth]{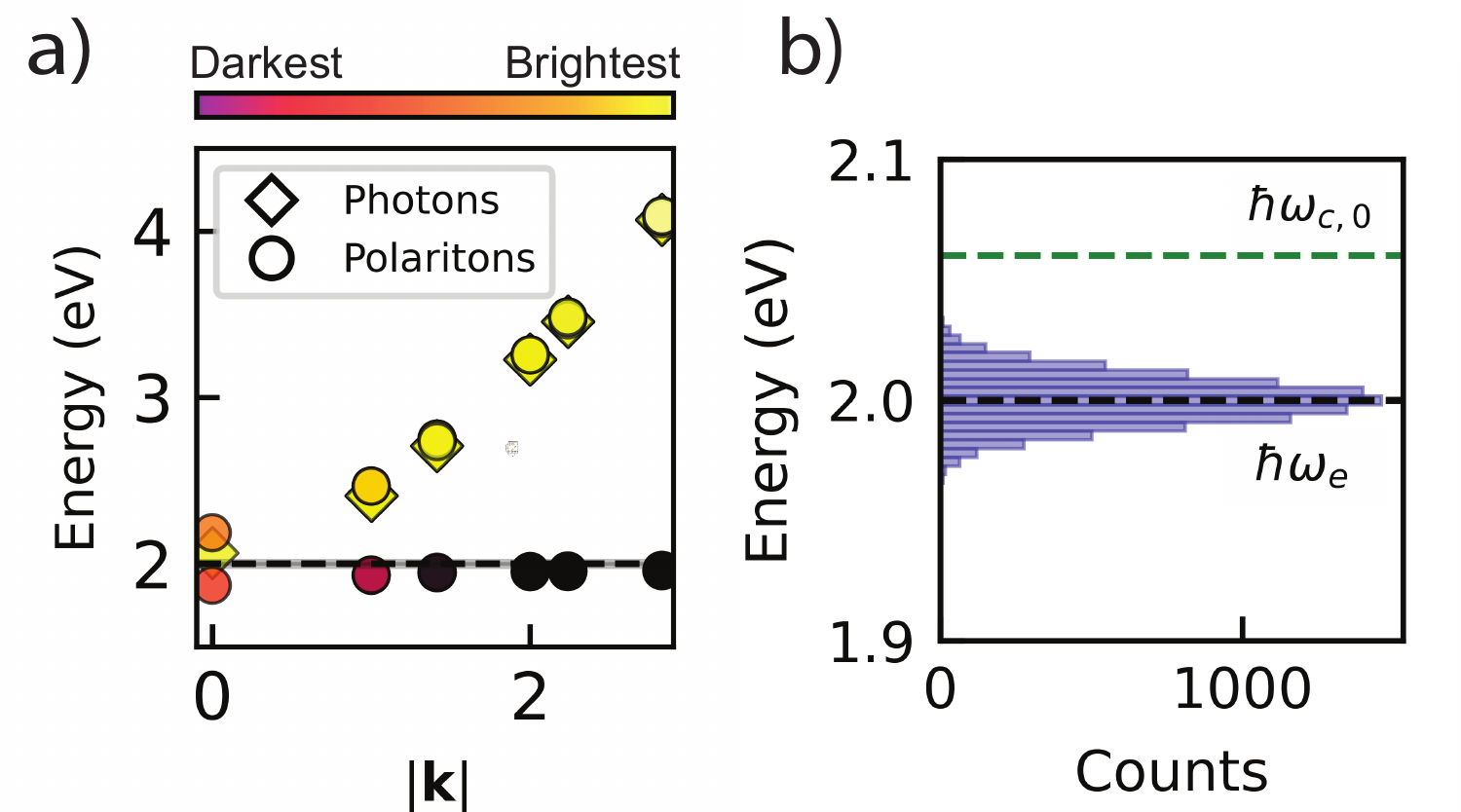}
    \vspace{-15pt}
    \caption{\textbf{Energy dispersion and distribution for a 3D cavity system with 10,201 molecules at the antinode (\( N_x = N_y = 101, N_z = 1 \)) and the cavity modes slightly blueshifted from the average molecular energy.} (a) Energy spectrum as a function of total wavevector $|\mathbf{k}|$. For each of the upper and lower polariton bands (circles), only the 25 lowest energy states are shown, which include near-degeneracies at equivalent wavevector magnitudes.
    The wavevector assigned to each state corresponds to the photonic mode with highest fraction (i.e., squared overlap). 
    The color scale represents total photonic fraction—yellow being the highest. 
    For reference, the uncoupled photonic modes (diamonds) and the average molecular exciton energy (black dotted line) are shown. 
    (b) The histogram depicts the distribution of molecular exciton energies sampled from a normal distribution with a mean of \( \hbar \omega_e = 2 \) eV and a standard deviation of \( \sigma_e = 0.01\) eV. The green dotted line marks the energy of the first photonic mode at \( \hbar \omega_{c,0} = 2.06 \) eV.}
    \label{dispersion}
\end{figure}
With the above formulation, we can choose the following parameters when constructing our systems. The average gap between the molecular ground state and its excited states is chosen to be $\hbar \omega_e = 2$ eV, with a disorder of $\sigma_e=0.01$ eV. We choose $L_z = 300$ nm and $\epsilon = 1$, which gives the lowest energy photonic state as $\hbar \omega_{c,0} = \hbar c\pi /(\sqrt{\epsilon} L_z) \approx 2.06$ eV. When placing the molecules, we take the fixed spacing between any two molecules to be $a_x = a_y = a_z = 10$ nm~\cite{Ribeiro2022}. Therefore, for a system with $N_x$ and $N_y$ molecules along the $x$ and $y$ axes, the wavevectors $\mathbf{k}$ then become $(2\pi m_x /[(N_x-1) a_x], 2\pi m_y/[(N_y-1) a_y])$, where $m_w = -(N_w-1) / 2, -(N_w-1) / 2 +1 , ... , (N_w-1) / 2 $ for $w = x, y$. This construction always ensures that, regardless of the system size, the largest wavevector magnitude remains constant because of the uniform spacing along the $x$ and $y$ axes. 
It follows that the energy range of the photonic modes does not change with system size.
Also, we choose $\mu$ such that the Rabi splitting is consistently set to a fixed value of \(0.2\) eV across all calculations.
By fixing these energy scales, we are able to isolate the effects of the parameters of interest, i.e., system size and disorder. In Figure \ref{dispersion}, we show the energy dispersion plot for one of the systems in our calculations.
Similar to 1D multimode models~\cite{Engelhardt2023}, the blueshift of the cavity relative to the molecules results in most eigenstates being essentially fully dark or fully bright, with a small number of eigenstates above and below the molecular energy having intermediate photon fraction.

\begin{figure}
    \centering
    \includegraphics[width = 1.0\textwidth]{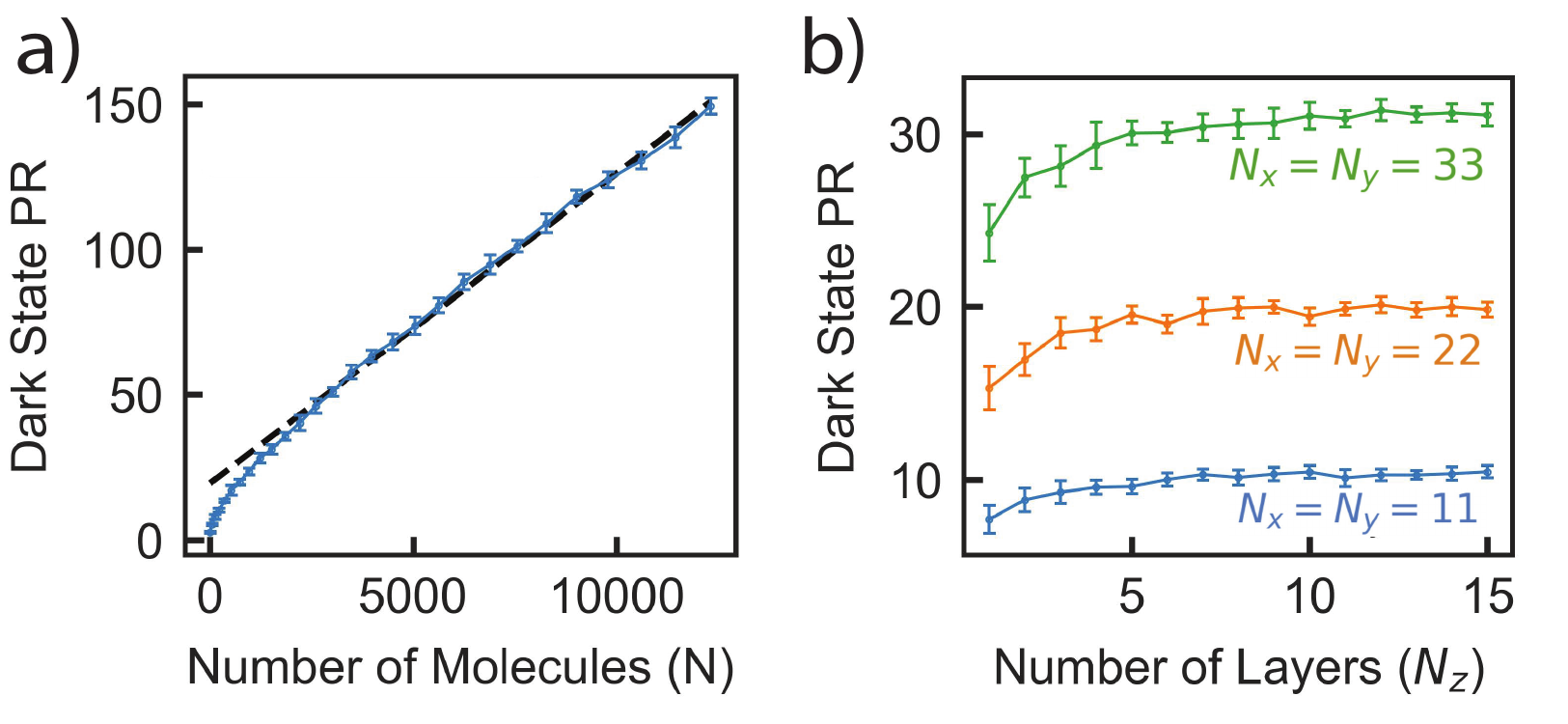}
    \caption{\textbf{The effect of increasing the number of molecules on dark-state delocalization.} (a) Changes in dark-state PR as more molecules are placed in a single layer in the \textit{xy} plane at the antinode of the cavity. A linear fit highlights the trend, with a slope approximately 0.0107, showing how PR increases with system size. (b) Changes in dark-state PR when more layers of molecules are placed in the cavity. Results demonstrate that PR remains relatively constant as the number of layers increases, across three distinct calculations with different values of fixed $N_x$ and $N_y$. The intermolecular distances ($a_x=a_y=10$ nm) and the Rabi splitting ($0.2$ eV) are kept constant throughout all calculations.}
    \label{fig:num_mol}
\end{figure}

\subsection{Increasing number of molecules}
Our first simulation focuses on a single layer of molecules placed at the antinode of the cavity, where the field amplitude, and thus the light-matter coupling, of the photonic modes is the largest. As shown in Figure \ref{fig:num_mol}, each data point corresponds to a square molecular configuration, with the system containing an equal number of molecules, $N_x = N_y$, along the $x$ and $y$ dimensions. As previously mentioned, the lattice maintains a fixed spacing of 10 nm, which increases the cavity length along the $x$ and $y$ dimensions as more molecules are added. In Figure \ref{fig:num_mol}a, we observe a linear trend for the dark-state PR after the number of molecules is larger than 2000. This indicates that cavity eigenstates tend to delocalize into more molecular states as the number of available molecular and photonic states increases with the system size. 
The linear scaling also suggests that the dark states include extended states. 
In contrast, the dark states in single-mode~\cite{Botzung2020, Du2022} and 1D multimode models~\cite{Ribeiro2022} are ``semilocalized''~\cite{Botzung2020} across a small number of molecules which does not scale with system size.

\begin{figure}
    \centering
    \includegraphics[width = 1.0\textwidth]{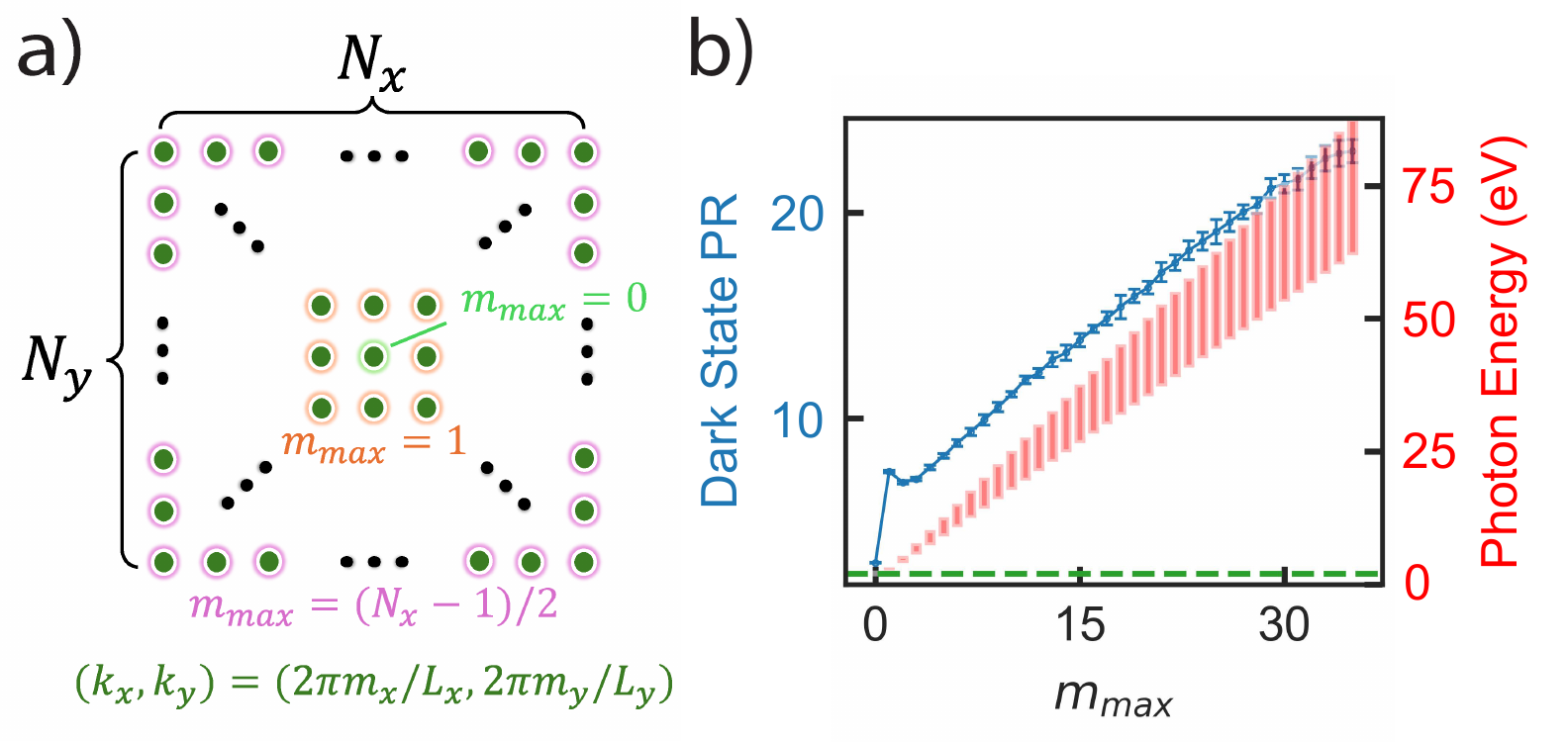}
    \caption{\textbf{Dark-state delocalization induced by photon modes of different energy for a 3D cavity system with 5041 molecules at the antinode ($ N_x = N_y = 71, N_z = 1 $).}   
    (a) Visualization of photon modes arranged according to in-plane wavevectors $(k_x, k_y) = (2 \pi m_x / L_x, 2 \pi m_y / L_y)$ and sorted into ``shells" corresponding respectively to different  $m_{\text{max}}=\max\{|m_x|, |m_y|\}$ and, thus, different energy ranges (see b).  
    Illustrated are shells $m_{\text{max}} = 0$ (green outline), $m_{\text{max}} = 1$ (orange outline), and $m_{\text{max}} = (N_x - 1) /2$ (pink outline).
    For each wavevector (green dot), there are two photon modes, corresponding to the two possible polarizations.
    The total number of photon modes in shell $m_{\text{max}}$ is 2 if $m_{\text{max}} = 0$ or $16 m_{\text{max}}$ if $m_{\text{max}} > 0$.
    (b) Dark-state PR (blue line with error bars) when molecules are coupled to individual shells of photon modes with different $m_{\text{max}}$. 
    The energy range of the photon modes for each shell $m_{\text{max}}$ is indicated by red bars, and the average molecular energy is represented by a green dotted line. 
    Higher $m_{\text{max}}$ corresponds to higher photon energies. 
    For most values of $m_{max}$, the photon energy is well above that of the molecules, suggesting that high-energy photonic modes collectively contribute to dark-state delocalization.}
    \label{fig:high-e}
\end{figure}

Notably, due to computational constraints, our largest simulation includes only 111 molecules along each of the $x$ and $y$ dimensions, with only a few photonic states being energetically close to the molecular states. 
To investigate how photonic modes with different energy influence the delocalization of dark states, we first partition the modes based on energy.
Specifically, as shown in Figure \ref{fig:high-e}a, we arrange the photonic modes according to their in-plane wavevector $(k_x, k_y) = (2 \pi m_x / L_x, 2 \pi m_y / L_y)$ and divide the modes into ``shells" with different $m_{\text{max}} = \max\{|m_x|, |m_y|\}$.
As the shell index increases, so does the average energy of the contained photonic modes (Figure \ref{fig:high-e}b, red bars). 
Then, for each shell, we carry out a calculation where we couple its photonic modes (ignoring those of other shells) to a single layer of molecules at the antinode of the cavity and calculate the dark-state PR. 
As shown in Figure \ref{fig:high-e}b, modes with higher $m_{max}$, whose photon energies lie further above the molecular energy, lead to greater dark-state PR.
These calculations are highly suggestive that for the actual system where the molecules are coupled to all photon modes, the increase in dark-state PR with the number of molecules (Figure \ref{fig:num_mol}a) might result from a collective effect induced by higher-energy photonic modes.

\subsection{Increasing number of layers}
This section examines the effect on dark-state delocalization when additional molecules are introduced along the $z$-axis of the cavity, while keeping the Rabi splitting constant. In this case, since the photonic modes considered depend only on the number and arrangement of molecules in the $xy$ plane, adding more molecular layers does not alter the photonic modes, despite the increase in the total number of molecules. Here, we start with three different square configurations of molecules of varying sizes, then gradually add more layers into the cavity as described in the previous section. Figure \ref{fig:num_mol}b shows only modest increases in dark-state PR with the addition of more molecules along the longitudinal ($z$-axis) of the optical cavity, particularly beyond 5 layers. This result reflects the fact that the molecules closer to the cavity mirrors feel a lower electric field intensity and hence have a smaller coupling to the cavity.  

\begin{figure}
    \centering
    \includegraphics[width = 1.0\textwidth]{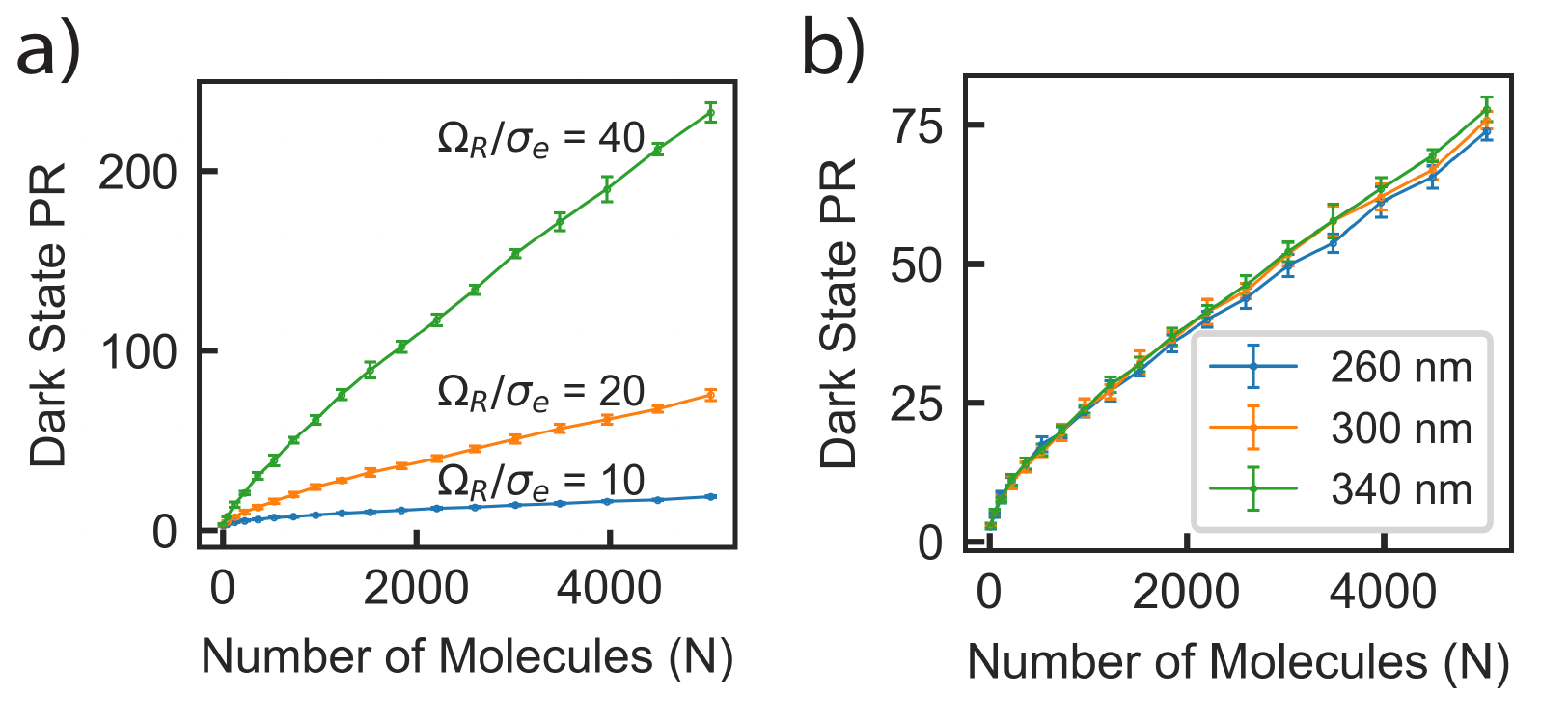}
    \caption{\textbf{Influence of molecular disorder and cavity length on the delocalization of dark states for a single layer of molecules placed at the cavity antinode.} (a) Dark-state PR versus number of molecules for various values of molecular disorder \(\sigma_e\) relative to Rabi splitting \(\Omega_R\).
    All curves become linear with varying slopes after 2000 molecules. 
    Higher \(\Omega_R / \sigma_e\) corresponds to faster scaling of dark-state delocalization with the number of molecules. 
    (b) Dark-state PR versus number of molecules for cavity lengths of 260, 300, and 340 nm, for which the lowest photon mode energies are \(\hbar \omega_{c,0} = 2.38\), \(2.06\), and \(1.82\) eV, respectively.}
    \label{fig:var}
\end{figure}

\subsection{Increasing molecular disorder}
In addition to examining the effect of molecular arrangement, we explore the relationship between the magnitude of molecular disorder and dark-state delocalization. 
Previous work~\cite{Ribeiro2022} shows that increased molecular disorder $\sigma_e$, relative to the Rabi splitting $\Omega_R$, leads to a smaller dark-state PR in a 1D multimode model. This relationship remains consistent in our experimentally representative model, as illustrated in Figure \ref{fig:var}a. 
In addition, the dark-state delocalization increases faster with the total number of molecules at higher $\Omega_R / \sigma_e$. 
Overall, smaller disorder, relative to the Rabi splitting, promotes greater delocalization among molecular states. Thus, for researchers interested in tuning dark-state delocalization in experiments, manipulating the Rabi splitting relative to the inhomogeneous broadening of the molecular absorption peak presents a viable approach.

\subsection{Changing the Longitudinal Length of Cavities}
Besides Rabi splitting, the length of the optical cavity is also a tunable parameter in experiments. As the cavity length $L_z$ increases, the lowest photonic energy $\omega_{c,\textbf{}{0}}$ decreases. To investigate its effect on dark-state delocalization, we make slight adjustments to the cavity length to ensure that the assumption of considering only the lowest photon band remains valid, keeping the lowest photonic energy close to the average molecular exciton energy. In Figure \ref{fig:var}b, although the three curves appear similar, a general trend indicates an increase in dark-state delocalization as the cavity length increases, suggesting that lowering $\omega_{c,\textbf{}{0}}$ could enhance coupling between light and dark states. However, it is worth noting that this change is less pronounced than the effect of $\Omega_R/\sigma_e$ discussed in the previous section. For researchers interested in exploring the effect of more significant variations in cavity length, it will be crucial to account for the higher-energy photon bands, an effect that is beyond the scope of the current study.

\section{Conclusions}

In this work, we study the delocalization of dark states accounting for the 3D geometry of molecules in a Fabry-P\'{e}rot cavity.
As in typical experiments, our model incorporates a disordered ensemble of molecules in a multidimensional arrangement and multiple photon modes which propagate freely in two dimensions and include all possible polarizations. 
Throughout all calculations, we keep intermolecular separations and (collective) Rabi splitting fixed.
We reveal a linear dependence of dark-state delocalization on the number of molecules along the $x$ and $y$ directions within the cavity. 
However, increasing the number of molecules along the $z$-axis leads to a relatively minimal change in delocalization. 
The linear scaling is in stark contrast to what is found in single-mode~\cite{Botzung2020, Du2022} and 1D multimode models~\cite{Ribeiro2022}, where the delocalization of the dark states instead saturates with number of molecules.
Similar to these simpler models, though, our calculations suggest that a higher ratio of Rabi splitting to molecular energy disorder, along with a wider cavity, enhance dark-state delocalization. 

Overall, our results encourage more experimental efforts to validate our theories and advocate more towards constructing realistic 3D cavity models to explain the experiments.
In this work, we consider only the lowest band of photonic modes, which corresponds to the lowest wavevector along the confinement axis ($z$).
Going beyond this approximation is expected to be important for understanding typical experiments of modified reaction kinetics, where the molecules are strongly coupled to the higher bands of photonic modes. 
Photonic modes in higher Brillouin zones should also be considered in future studies, given our result (Figure~\ref{fig:high-e}) suggesting that high-energy photonic modes might contribute significantly to the delocalization of dark states. 
Including these currently neglected modes in our calculations is anticipated to further delocalize the dark states and leave our main conclusions unchanged. 
Regarding the molecules, future work should more carefully explore the effect of positional disorder, which is relevant to understanding solution-phase chemistry under strong coupling.
Specifically, while we have found that our results are robust to large positional disorder, such conditions break translation symmetry and hence should (at least in principle) be accompanied by the entire continuum of photon modes, not (as is the case here) just the modes corresponding to a discrete set of reciprocal lattice vectors.
It could also be interesting to explore the effect of non-static disorder, which has been explored in single-mode models~\cite{Sommer2021, Chen2022, Zhou2023}.
For simplicity, one might start with disorder that evolves as white noise~\cite{Haken1972, Haken1973, Chen2022, Zhou2023}.
To properly understand how the delocalization of dark states affects chemistry inside an optical cavity, it will be essential to make comparisons with molecules coupled to free-space photons~\cite{Vurgaftman2022}, which can also feature polaritons and dark states. 

\section{Acknowledgements}

We dedicate this study to Naomi Halas and Peter Nordlander, whose leadership in the Plasmonics and Nanophotonics community has inspired future generations of scientists to creatively partake in its interdisciplinary endeavors.
K.S. acknowledges funding support from an Summer 2020 Undergraduate Research Award from the Division of Physical Sciences at UCSD. 
M.D. and J.Y.-Z. were supported by the U.S. Department of Energy, Office of Science, Basic Energy Sciences, CPIMS Program under Early Career Research Program award DE-SC0019188.

\providecommand{\latin}[1]{#1}
\makeatletter
\providecommand{\doi}
  {\begingroup\let\do\@makeother\dospecials
  \catcode`\{=1 \catcode`\}=2 \doi@aux}
\providecommand{\doi@aux}[1]{\endgroup\texttt{#1}}
\makeatother
\providecommand*\mcitethebibliography{\thebibliography}
\csname @ifundefined\endcsname{endmcitethebibliography}
  {\let\endmcitethebibliography\endthebibliography}{}

\end{document}